\documentclass[english]{article}
\usepackage[T1]{fontenc}
\usepackage[latin9]{inputenc}
\usepackage{amsmath}
\usepackage{amsthm}
\usepackage{amssymb}

\makeatletter

\providecommand{\tabularnewline}{\\}

\numberwithin{equation}{section}
\numberwithin{figure}{section}
\newcommand{\lyxaddress}[1]{
	\par {\raggedright #1
	\vspace{1.4em}
	\noindent\par}
}
\theoremstyle{plain}
\newtheorem*{thm*}{\protect\theoremname}
\theoremstyle{plain}
\newtheorem{thm}{\protect\theoremname}

\usepackage{graphicx}

\makeatother

\usepackage{babel}
\providecommand{\theoremname}{Theorem}

\begin{document}
\title{Gauge Theories With Infinite Multiplets of Fermions}
\author{S. G. Rajeev}
\maketitle

\lyxaddress{Department of Physics and Astronomy, University of Rochester, Rochester
NY 14627}
\begin{abstract}
We study the coupling constant renormalization of gauge theories with an infinite multiplet of fermions, using the zeta function method to make sense of the infinite sums over fermions. If the gauge group $K$ is the maximal compact subgroup of a simple non-compact group $G$, such infinite multiplets can arise naturally, as reductions of discrete series unitary representations of $G$. The example $K=U(1)\subset SU(1,1)=G$ will be studied in detail. Surprisingly,  there are abelian gauge theories which are asymptotically free; and others that are UV finite.
\end{abstract}

\section{Introduction}

The signs the of the first two coefficients (one and two loop) of
the beta function of a gauge theory are of great significance\cite{QFT}.
The discovery that the leading coefficient is negative\cite{AsympFreedom}
(asymptotic freedom), for QCD with a small enough number of quark
flavors, explained scaling in Deep Inelastic Scattering.The second
coefficient\cite{CaswellJones} is also negative for QCD with a small
enough number of quark flavors. But there is a range of values of
$N_{f}$ where it is positive while preserving asymptotic freedom.
This leads to a non-trivial IR fixed point often called the Banks-Zaks
fixed point\cite{BanksZaks}. \footnote{The coefficients of higher order terms, while of great interest in
precision tests of gauge theories, do not change the qualitative nature
of the fixed points. Moreover, they are ``scheme dependent''; i.e.,
are affected by a redefinition of the coupling constant $\alpha\mapsto f(\alpha)$.}

By contrast, both these coefficients of the beta function are positive
for QED. With just one Dirac fermion of unit charge (say the electron)
the beta function is\cite{QEDBetaFn} 

\[
\beta(\alpha)=\frac{2}{3}\frac{\alpha}{\pi}+\frac{1}{2}\left(\frac{\alpha}{\pi}\right)^{2}+\mathrm{O}(\alpha^{3})
\]

A physical interpretation is that vacuum polarization leads to a point
charge being screened by virtual anti-particles; so as we get closer
to the charge, its strength will increase. By contrast, in non-abelian
gauge theories the ``paramagnetic'' contribution from virtual gluons
leads to anti-screening\cite{Polyakov}.

To this order, only diagrams with a single fermion loop contribute.
So if we have a multiplet of Dirac fermions with charges $e_{\nu}$
we would have\footnote{At the next order some diagrams (with a single fermion loop) would
be modified a factor $\sum_{\nu}e_{\nu}^{6}$ while others (with two
fermion loops) would be modified by a factor $\sum_{\nu}e_{\nu}^{4}\sum_{\nu}e_{\nu}^{2}$.
Each fermion loop will have a factor $\sum_{\nu}e_{\nu}^{r}$ where
$r$ is the number of vertices. By Furry's theorem $r$has to be even
(see below).}

\begin{equation}
\beta(\alpha)=\frac{2}{3}\frac{\alpha}{\pi}\sum_{\nu}e_{\nu}^{2}+\frac{1}{2}\left(\frac{\alpha}{\pi}\right)^{2}\sum_{\nu}e_{\nu}^{4}+\mathrm{O}(\alpha^{3}),\quad\label{eq:RafaelRosner}
\end{equation}

Thus we cannot have asymptotic freedom or IR stable fixed points in
abelian gauge theories with finite multiplets of fermions.

In this paper we look into the possibility of an infinite multiplet
of charged fermions, for which sums such as $\sum_{\nu}e_{\nu}^{2}$
are divergent . Divergences occur often in Quantum Field Theory and
are given a meaning through regularization and renormalization. For
example, using zeta function regularization. 

The basic mathematical model is the Riemann zeta function $\zeta(s)=\sum_{n=1}^{\infty}n^{-s}$
. It converges when $\mathrm{Re}\ s>1$. The function can be extended
to the whole of the complex plane by analytic continuation. Other
than a simple pole at $s=1$ it is regular. The values of at negative
integer values of $s$ are well-known\cite{Apostol}:
\[
\zeta(0)=-\frac{1}{2},\quad\zeta(-1)=-\frac{1}{12},\quad\zeta(-2)=0,\quad\zeta(-3)=\frac{1}{120},\quad\zeta(-4)=0,\cdots
\]

This allows us to give a meaning to sums of powers of natural numbers:
\[
\sum_{n=1}n^{0}=-\frac{1}{2},\quad\sum_{n=1}n=-\frac{1}{12},\quad\sum_{n=1}n^{2}=0,\quad\sum_{n=1}n^{3}=\frac{1}{120},\quad\sum_{n=1}n^{4}=0,\]

So, an infinite sum of positive numbers can have a ``renormalized''
value which is negative, zero or positive.

Can we get an asymptotically free abelian gauge theory by choosing
an infinite multiplet of charges $e_{\nu}$? We need an approximate
(``custodial'') global symmetry that preserves the relative ratios of these
charges: only an overall constant multiple will be renormalized. One
way to do this is to have a global symmetry under a non-compact group
$G$ (such as $SU(1,1)$ ) of which the gauge group $K=U(1)$ is a
maximal compact subgroup. A unitary representation of $G$ is necessarily
infinite dimensional. It will decompose into an infinite sum of irreducible
representations (given by the charges $e_{\nu}$) of $K$ . The symmetry
under $G$ is broken by the gauge couplings. In the example of a discrete
series representation\cite{Bargmann} of $SU(1,1)$ the charges form
an arithmetic sequence

\[
e_{\nu}=k+\nu,\quad\nu=0,1,\cdots,\quad k>0.
\]

We will see that the resulting abelian gauge theory is 
\begin{itemize}
\item asymptotically free when $0<k<\frac{1}{2}$ or if $k>1$; with a non-trivial
IR stable fixed point (``Banks-Zaks'') for $0<k<\frac{1}{2}$.
\item asymptotically safe when $\frac{1}{2}<k<1$ (i.e., has a non-trivial
UV stable fixed point)
\item UV finite to all orders of perturbation theory when $k=\frac{1}{2}$
or $k=1$
\end{itemize}
The low-energy behavior, such as the symmetry of the vacuum and the
spectrum of bound states of these theories lie beyond the reach of
perturbation theory.

It is also possible to extend these ideas to some non-abelian gauge
groups $K$. An elegant choice is the maximal compact subgroup $K\subset G$
of a non-compact Lie group $G$ admitting a discrete series representation.
All the essential information needed to calculate the beta function
is contained in the character formula of Harish-Chandra, which generalizes
the Weyl character formula of compact semi-simple group representations.
The calculations involved are substantially more complicated. So we
will only give an outline of the ideas. We hope to return to this
in a later publication.

\section{The Discrete Series $\underline{SU}(1,1)$}

We will now give a self-contained derivation of the discrete series
representations\cite{Bargmann} of $SU(1,1)$. There is nothing new
in our description here. The argument is a small modification of the
standard angular momentum theory of quantum mechanics\cite{Georgi}
. The original paper\cite{Bargmann} as well as later expositions\cite{Knapp,Varadarajan}
deal with much more general cases and are in a notation unfamiliar
to physicists.

$SU(1,1)$ is the group of complex $2\times2$ matrices satisfying
\[
\det g=1,\quad g\sigma_{3}g^{\dagger}=\sigma_{3},\quad\sigma_{3}=\left(\begin{array}{cc}
1 & 0\\
0 & -1
\end{array}\right)
\]

while its Lie algebra \textbf{$\underline{SU}(1,1)$} is the real
vector space of matrices satisfying
\[
\mathrm{tr}\gamma=0,\quad\gamma\sigma_{3}+\sigma_{3}\gamma^{\dagger}=0.
\]

A basis is 
\[
e_{0}=\frac{i}{2}\sigma_{3},\quad e_{1}=\frac{1}{2}\left(\begin{array}{cc}
0 & 1\\
1 & 0
\end{array}\right),\quad e_{2}=\frac{1}{2}\left(\begin{array}{cc}
0 & -i\\
i & 0
\end{array}\right)
\]
satisfying the commutation relations 
\[
[e_{0},e_{1}]=-e_{2},\quad[e_{0},e_{2}]=e_{1},\quad[e_{1},e_{2}]=e_{0}.
\]

The maximal compact sub-algebra $\underline{U}(1)$ has basis $e_{0}$.
The group elements are $g=e^{\xi_{0}e_{0}+\xi_{1}e_{1}+\xi_{2}e_{2}}\in SU(1,1)$
for real $\xi_{0},\xi_{1},\xi_{2}$. 

\subsection{The Representation $D_{k}$ of the Lie Algebra $\underline{SU}(1,1)$}

So, a representation of $\underline{SU}(1,1)$ consists of operators
$\hat{e}_{0},\hat{e}_{1},\hat{e}_{2}$ satisfying 
\[
[\hat{e}_{0},\hat{e}_{1}]=-\hat{e}_{2},\quad[\hat{e}_{0},\hat{e}_{2}]=\hat{e}_{1},\quad[\hat{e}_{1},\hat{e}_{2}]=\hat{e}_{0}
\]

So in a unitary representation of the group, $\hat{g}=e^{\xi_{0}\hat{e}_{0}+\xi_{1}\hat{e}_{1}+\xi_{2}\hat{e}_{2}}$
must be unitary for real $\xi_{0},\xi_{1},\xi_{2}$. This requires
\[
\hat{e}_{0}^{\dagger}=-\hat{e}_{0},\quad\hat{e}_{1}^{\dagger}=-\hat{e}_{1},\quad\hat{e}_{2}=-\hat{e}_{2}.
\]

Our $2$ dimensional representation satisfies the first of these conditions
but not the other two; it is not unitary. Indeed all unitary representations
are infinite dimensional.

It is useful to define\footnote{Our notation is related to Bargmann's original one by the table

\begin{tabular}{|c|c|}
\hline 
Us & Bargmann\tabularnewline
\hline 
\hline 
$J_{-}$ & $F$\tabularnewline
\hline 
$J_{+}$ & $G$\tabularnewline
\hline 
$J_{0}$ & $-H_{0}$\tabularnewline
\hline 
$\frac{J_{+}J_{-}+J_{-}J_{+}}{2}-J_{0}^{2}$ & $Q$\tabularnewline
\hline 
$-k(k-1)$ & $q$\tabularnewline
\hline 
$k$ & $-\lambda$\tabularnewline
\hline 
$\mid\nu\rangle$ & $f_{-k-\nu}$\tabularnewline
\hline 
$\mid0\rangle$ & $g$\tabularnewline
\hline 
\end{tabular}}
\[
J_{0}=-i\hat{e}_{0},\quad J_{\pm}=-i\left(\frac{\hat{e}_{1}\pm i\hat{e}_{2}}{\sqrt{2}}\right)
\]

so that in a unitary representation

\[
J_{0}^{\dagger}=J_{0},\quad J_{-}^{\dagger}=J_{+}
\]

and 
\[
[J_{0},J_{-}]=-J_{-},\quad[J_{0},J_{+}]=J_{+},\quad[J_{-},J_{+}]=J_{0}.
\]

The angular momentum operators (of $SU(2)$ ) are similar, except
that the last relation has the opposite sign.

We can choose an orthonormal basis in which $J_{0}$ is diagonal.
Then $J_{-}$ lowers the eigenvalue of $J_{0}$ by one unit while
$J_{+}$ raises it by the same amount. 

We can look for a representation built on a lowest weight state 
\[
J_{-}\mid0\rangle=0,\quad J_{0}\mid0\rangle=k\mid0\rangle
\]

In an irreducible lowest weight representation, the remaining basis
elements are obtained by repeated action of $J_{+}$ on this state.

The condition $[J_{0},J_{+}]=J_{+}$ suggests the ansatz
\[
J_{+}\mid\nu\rangle=f(\nu)\mid\nu+1\rangle,\quad J_{0}\mid\nu\rangle=(k+\nu)\mid\nu\rangle,\quad\nu=0,1,2,\cdots
\]

Unitarity will imply $\langle\nu'\mid J_{-}\mid\nu\rangle=\langle\nu\mid J_{+}\mid\nu'\rangle^{*}=\delta_{\nu,\nu'+1}f^{*}(\nu')$
or 
\[
J_{-}\mid\nu\rangle=f^{*}(\nu-1)\mid\nu-1\rangle.
\]
This gives $[J_{0},J_{-}]=-J_{-}$ immediately. Also, $J_{-}\mid0\rangle=0$
implies 
\[
f(-1)=0.
\]
.

Finally $[J_{-},J_{+}]=J_{0}$ gives the condition 
\[
\mid f(\nu)\mid^{2}-\mid f(\nu-1)\mid^{2}=(k+\nu)
\]

This difference equation can be solved using the initial condition
$f(-1)=0$ :
\[
\mid f(\nu)\mid^{2}=\frac{1}{2}(\nu+1)(\nu+2k)
\]

Up to a phase that can be removed by redefining the states $\mid\nu\rangle$
we get 
\[
f(\nu)=\sqrt{\frac{1}{2}(\nu+1)(\nu+2k)},\quad\nu=0,1,\cdots
\]

This is all very similar to the usual construction of lowest weight
unitary representations of $\underline{SU}(2)$; except that $f(\nu)\neq0$
for all $\nu=0,1,2,\cdots$.This is an infinite dimensional representation:
there is no highest weight state.

In summary, we have a unitary representation $D_{k}$ 
\[
J_{0}\mid\nu\rangle=(k+\nu)\mid\nu\rangle,\quad\nu=0,1,2,\cdots
\]
\[
J_{-}\mid\nu\rangle=\sqrt{\frac{1}{2}\nu(\nu+2k-1)}\mid\nu-1\rangle
\]
\[
J_{+}\mid\nu\rangle=\sqrt{\frac{1}{2}(\nu+1)(\nu+2k)}\mid\nu+1\rangle
\]

Since $\langle0\mid J_{-}J_{+}\mid0\rangle=k$ we have the condition
\[
k>0.
\]
 For historical reasons these representations are called the ``Discrete
Series''.  We emphasize that, despite this name, the range of allowed
values of $k$ is continuous: for any postive $k$ we have a representation
$D_{k}$ of the Lie algebra.

\subsection{Representation $D_{k}$ of $SU(1,1)$ and its Covering Groups}

The group $SU(1,1)$ is homotopic to its maximal compact subgroup
$U(1)$. This is the subgroup of elements of the type $\exp\left(\xi_{0}e_{0}\right)$
with $e_{0}=\frac{i}{2}\sigma_{3}$ and $\xi_{0}\in\mathbb{R}$. Note
that 
\[
\exp\left(4\pi e_{0}\right)=1.
\]

In any representation of $SU(1,1)$ we must have 
\[
\exp(4\pi\hat{e}_{0})=\exp\left(4\pi iJ_{0}\right)=1.
\]
Thus, for our lowest weight unitary representation $D_{k}$ of $\underline{SU}(1,1)$
to exponentiate to a representation of $SU(1,1)$ , $k$ must be an
integer or half-integer. This is why it was called the discrete series
historically.

Now, recall that $\underline{SU}(1,1)$ is the Lie algebra of many
Lie groups which are related to each other by coverings. If $k$ is
an integer, $D_{k}$ exponentiates to a unitary irreducible representation
of $SO(1,2)$. If $k$ is a half-integer it gives a representation
of the double cover $SU(1,1)$. A rational $k$ will exponentiate
to some finite cover of $SU(1,1)$. If $k$ is irrational, $D_{k}$
exponentiates to a representation of the universal covering group
$\widetilde{SU}(1,1)$.

\subsection{The Oscillator Representations $D_{\frac{1}{4}}$ and $D_{\frac{3}{4}}$}

Of special interest are two representations that arise from the harmonic
oscillator:
\[
J_{-}=\frac{a^{2}}{2\sqrt{2}},\quad J_{+}=\frac{a^{\dagger2}}{2\sqrt{2}},\quad J_{0}=\frac{1}{2}\left(a^{\dagger}a+\frac{1}{2}\right)
\]

where
\[
a\mid n)=\sqrt{n}\mid n-1),\quad a^{\dagger}\mid n)=\sqrt{n+1}\mid n+1)
\]

satisfying 

\[
[a,a^{\dagger}]=1.
\]

Then

\[
J_{-}\mid n)=\frac{1}{2\sqrt{2}}\sqrt{n(n-1)}|n-2),\quad J_{+}\mid n)=\frac{1}{2\sqrt{2}}\sqrt{(n+1)(n+2)}\mid n+2)
\]

\[
J_{0}\mid n)=\frac{1}{2}\left(n+\frac{1}{2}\right)\mid n)
\]

When we compare to the above formulas, we can see this corresponds
to a sum of two irreducible representations. In one case we have even
occupation numbers 

\[
\mid\nu\rangle=\mid2\nu)
\]

and odd in the other:

\[
\mid\nu\rangle=\mid2\nu+1).
\]

Comparing the lowest weights 

\[
J_{0}\mid0)=\frac{1}{4}\mid0),\quad J_{0}\mid1)=\frac{3}{4}\mid0)
\]

we see that the even representation has $k=\frac{1}{4}$ and the odd
one has $k=\frac{3}{4}$

These exponentiate to a representation of the double cover $Mp(1,1)$
(called the ``metaplectic group'' ) of $SU(1,1)$. 

Other representations can be built from higher dimensional oscillators
by forming rotation invariant combinations of bi-linears in $a,a^{\dagger}$.
But we won't pursue these constructions here.

\subsection{The Character Function of $D_{k}$}

Given a finite dimensional representation , $\rho:G\to U(\mathcal{V})$
of a group, the character function $\chi_{\rho}:G\to\mathbb{C}$ is
the trace $\chi(g)=\mathrm{tr}\rho(g)$. For an infinite dimensional
representation, this trace may not be a convergent sum. Even so, the
character can exist as a distribution or generalized function. Being
invariant under conjugation, it can be reduced to a function on a
Cartan subgroup (a subgroup of simultaneously diagonalizable elements).
For $G=SU(1,1)$ there two conjugacy classes of Cartan subgroups:
$\left(\begin{array}{cc}
e^{i\theta} & 0\\
0 & e^{-i\theta}
\end{array}\right),0\leq\theta<2\pi$ and $\left(\begin{array}{cc}
\cosh\zeta & \sinh\zeta\\
\sin\zeta & \cosh\zeta
\end{array}\right),\zeta>0$ .

The first of these Cartan subgroups is generated by $J_{0}$ . Restricted
to this, the character of the discrete series with parameter $k$
is 

\[
\mathrm{tr}e^{i\theta J_{0}}=\sum_{\nu=0}^{\infty}e^{i\theta(k+\nu)}
\]

The sum does not converge: the character is a distribution rather
than a function of $\theta$. If we allow $\theta$ to have a small
positive imaginary part, the sum will converge. It is more convenient
to define $\theta=i\tau$ and set 

\[
\chi_{k}(\tau)\equiv\mathrm{tr}e^{-\tau J_{0}}=\sum_{\nu=0}^{\infty}e^{-\tau(k+\nu)}
\]

This sum converges for $\mathrm{Re}\ \tau>0$:

\[
\chi_{k}(\tau)=\frac{e^{-k\tau}}{1-e^{-\tau}}
\]

The quantity on the r.h.s. has an analytic continuation to the whole
$\tau-$plane, with a simple pole at $\tau=0$. In other words, the
character is a generalized function on the unit circle $U(1)\subset SU(1,1)$
, which is the boundary value of an analytic function in the interior
of the unit disc.

The Casimir invariants of a finite dimensional representation are
traces of powers of the Lie algebra representatives

\[
z(r,k)=\mathrm{tr}J_{0}^{r}
\]
For a finite dimensional representation, the character function is
the generating function of these Casimirs:

\[
\chi_{k}(\tau)=\sum_{r=0}^{\infty}\mathrm{tr}J_{0}^{r}\frac{(-\tau)^{r}}{r!}.
\]
For infinite dimensional representations such as $D_{k}$, the traces
$\mathrm{tr}J_{0}^{r}$ diverge. But, we can again give them a meaning
by analytic continuation.

One approach is to expand in a Laurent series 

\[
\chi_{k}(\tau)=\frac{1}{\tau}+\sum_{r=0}^{\infty}z(r,k)\frac{(-\tau)^{r}}{r!}
\]

By subtracting out the simple pole at $\tau=0$ , we get well-defined
answers for $z(r,k)$. Another, related, method is to define the zeta
function by the Mellin transform

\[
\zeta_{H}(s,k)=\frac{1}{\Gamma(s)}\int_{0}^{\infty}\chi_{k}(\tau)\tau^{s-1}d\tau,\quad\mathrm{Re}\ s>1
\]

In our case, we can evaluate this integral to get 

\[
\zeta_{H}(s,k)=\sum_{\nu=0}^{\infty}\frac{1}{(k+\nu)^{s}}
\]
This the Hurwitz zeta function\cite{Apostol}. It has a simple pole
at $s=1$ and is regular elsewhere.

Then we define 
\[
z(r,k)=\zeta_{H}(-r,k)
\]

Both methods give the same same answers, in terms of Bernoulli polynomials:

\[
z(0,k)=\frac{1}{2}-k
\]

\[
z(1,k)=\frac{1}{12}(-6k^{2}+6k-1)
\]

\[
z(2,k)=-\frac{1}{6}(k-1)k(2k-1)
\]

\[
z(3,k)=\frac{1}{120}(1-30k^{2}+60k^{3}-30k^{4})
\]

\[
z(4,k)=-\frac{1}{30}(k-1)k(2k-1)(-1-3k+3k^{2})
\]

\[
\cdots
\]

The Harish-Chandra gives a general formula for the character of a
discrete series representation. Expanding it in a series as above
allows us to extract Casimir invariants of the representation directly
for more general groups.

\section{The Beta Function of Abelian Gauge Theory}

Now we consider a set of mass-less Dirac fermions transforming under
the representation $D_{k}$ of an internal symmetry under the Lie
algebra $\underline{SU}(1,1)$ . Under the maximal compact sub-algebera
$\underline{U}(1)\subset\underline{SU}(1,1)$ we have an infinite
multiplet of charges 

\[
e_{\nu}=k+\nu,\quad\nu=0,1,\cdots
\]

We now couple these charges to an abelian gauge field. 

The two loop beta function is, the formula (\ref{eq:RafaelRosner})
becomes

\[
\beta(\alpha)=\frac{2}{3}\frac{\alpha}{\pi}z(2,k)+\frac{1}{2}\left(\frac{\alpha}{\pi}\right)^{2}z(4,k)+\mathrm{O}(\alpha^{3})
\]

If $z(r,k)$ are given a meaning through regularization as above,
we get 

\[
\beta(\alpha)=-\frac{1}{9}(k-1)k(2k-1)\left(\frac{\alpha}{\pi}\right)-\frac{1}{60}(k-1)k(2k-1)(3k^{2}-3k-1)\left(\frac{\alpha}{\pi}\right)^{2}+\mathrm{O}(\alpha^{3})
\]

\[
\equiv\beta_{1}\left(\frac{\alpha}{\pi}\right)+\beta_{2}\left(\frac{\alpha}{\pi}\right)^{2}+\mathrm{O}\left(\alpha^{3}\right)
\]

The coefficients of $\frac{\alpha}{\pi}$ and $\left(\frac{\alpha}{\pi}\right)^{2}$
are plotted in Fig. 4.1.

\begin{figure}[h]
\caption{The One and Two loop coefficients of the beta function} \centering 
\includegraphics[width=0.75\textwidth]{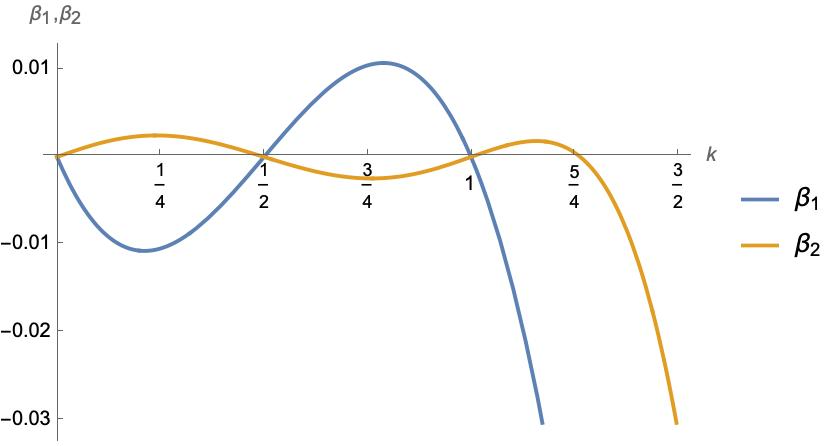} 
\label{fig:1}
\end{figure}

\subsection{Some Consequences}

So we have 
\begin{itemize}
\item asymptotic freedom for $0<k<\frac{1}{2}$ and $k>1$ (i.e., $\beta_{1}<0$)
\item For $0<k<\frac{1}{2}$ and for $1<k<\frac{1}{2}+\frac{\sqrt{21}}{6}\approx1.26$
there is a non-trivial IR stable fixed point (its location will depend
on higher order corrections which are scheme dependent)
\item For $\frac{1}{2}<k<\frac{1}{2}+\frac{\sqrt{21}}{6}\approx1.26$ we
have ``asymptotic safety'': at short distances, the coupling grows
from zero to a finite value of order one (whose value again depends
on the ``scheme'' chosen)
\end{itemize}
It should be emphasized that these conclusions are based on perturbation
theory. Without a non-perturbative method (e.g., lattice simulations)
we cannot be confident of their validity. But perhaps this gives some
encouragement to explore these phenomena using non-perturbative methods.

\subsection{When $k=\frac{1}{2}$ or $k=1$}

Let us recall Furry's theorem\cite{Furry } in abelian gauge theory. 
\begin{thm*}
Any Feynman diagram, with a sub-diagram consisting of a closed loop
of Dirac fermions with an odd number of vertices, is zero 
\end{thm*}
The reason is that a subdiagram with $r$ vertices will be proportional
to a trace over a product of Dirac matrices 

\[
\left(\sum_{\nu}e_{\nu}^{r}\right)\ \mathrm{tr}\gamma^{\mu_{1}}\cdots\gamma^{\mu_{r}},\quad r\ge1
\]

We now recall that there is a Dirac matrix $\gamma_{5}$ satisfying

\[
\gamma_{5}\gamma^{\mu}=-\gamma^{\mu}\gamma_{5},\quad\gamma_{5}^{2}=1
\]

So the trace is equal to its negative when $r$ is odd:

\[
\mathrm{tr}\gamma^{\mu_{1}}\cdots\gamma^{\mu_{r}}=\sum_{\nu}e_{\nu}^{r}\ \mathrm{tr}\gamma_{5}^{2}\gamma^{\mu_{1}}\cdots\gamma^{\mu_{r}}=\ \mathrm{tr}\gamma_{5}\gamma^{\mu_{1}}\cdots\gamma^{\mu_{r}}\gamma_{5}=\ (-1)^{r}\mathrm{tr}\gamma_{5}^{2}\gamma^{\mu_{1}}\cdots\gamma^{\mu_{2n+1}}.
\]

Thus only $z(r,k)=\sum_{\nu}e_{\nu}^{r}$ with $r$even can occur. 

We saw that for $k=\frac{1}{2},1$ these are zero:

\[
z(2r,k)=0,\quad r\geq1
\]

So the beta function of an abelian gauge theory with $k=\frac{1}{2},1$
is zero to all orders of perturbation theory.

Thus, abelian gauge theory the massless fermions in either of the
two discrete series with $k=\frac{1}{2},1$ is likely to be a conformal
field theory. This also apply to the limit case $k\to0^{+}$ (which
have non-zero values for $z(r,k)$, as we saw.) 

It would be very interesting to test this by non-perturbative methods
such as those of Ref.\cite{ConformalBootstrap}. 

\section{Non-Compact Simple Lie Groups and their Maximal Compact Subgroups}

A faithful unitary representation of a Lie group $G$ is a continuous
surjective map $\rho:G\to U(\mathcal{V})$ to the space of unitary
operators in a Hilbert space $\mathcal{V}$. Thus, the image $\rho(G)\subset U(\mathcal{H})$
is homeomorphic to $G$.

If $\mathcal{V}$ is finite dimensional, $U(\mathcal{V})$ is a compact
space. So $\rho(G)$ and hence $G$ itself, must be compact.Any faithful
representation of a non-compact Lie group is necessarily infinite
dimensional. 

\subsection{Discrete Series}

A particularly interesting class of unitary irreducible representations
(unirreps) are the discrete series. These are representations where
the matrix elements $\langle\psi\mid\rho(g)\mid\chi\rangle$ are square
integrable functions on $G$. The mathematics is deep, with connections
to number theory (e.g., Langlands program).

Not all non-compact Simple Lie groups have discrete series. Harish-Chandra\cite{DiscreteSeries}
found the criterion for when discrete series representations exist\cite{Knapp,Varadarajan}.
\begin{thm}
(Harish-Chandra) A linear semi-simple Lie group $G$ has Discrete
Series representations iff its rank is the same as that of its maximal
compact sub-group $K$
\end{thm}

Thus, $SO(1,2)$ has a discrete series \cite{Bargmann} but not $SO(1,3)$.
In fact $ SO(m,n)$ has discrete
series precisely when $mn$ is even. Also, $SU(m,n)$ has discrete
series for all $m,n\geq1$. The particular case $SU(2,3)$ is intriguing
as its maximal compact subgroup $S\left(U(2)\times U(3)\right)$ is
the gauge group of the standard model of particle physics. 

\subsection{Gauge Groups have to be compact}

Non-compact Lie groups cannot be gauge groups of Yang-Mills theories. This is well-known, but we give a reminder here of why.
That is because the pure Yang-Mills action can be written as 
\[
L_{YM}=-\frac{1}{4}g_{ab}F_{\mu\nu}^{a}F^{b\mu\nu}
\]
where $a,b=1,\cdots d$ label a basis in the Lie algebra $\underline{K}$
of the gauge group $K$ 
\[
[e_{a},e_{b}]=f_{ab}^{c}e_{c}
\]
For the action to be gauge invariant, the symmetric matrix $g_{ab}$
must be an invariant inner product on $\underline{K}$:
\[
f_{ab}^{d}g_{dc}+g_{ad}f_{bc}^{d}=0.
\]
 Also, to have a positive inner product in the quantum Hilbert space
we need $g_{ab}$ to be a positive matrix, in order that the space
of quantum states has a positive inner product\cite{NoGhost}. 

So at best we can gauge some sub-algebra $\underline{K}\subset\underline{G}$
which has a positive invariant inner product $g$.

Such a Lie algebra $\underline{K}$ with a positive invariant inner
product can exponentiate to a compact Lie group $K$. So we will say
that such Lie algebras of ``compact type''.\footnote{A subtle point is that a Lie algebra of compact type can exponentiate
to a non-compact Lie group; and the universal cover of a compact Lie
group need to be compact (think of $U(1)$ , whose universal cover
is $\mathbb{R}$ ). But this only happens for the abelian factors. }They are direct sums of compact simple Lie algebras and some abelian
Lie algebra . The most famous example is the standard model: $\underline{K}=\underline{U}(1)\oplus\underline{SU}(2)\oplus\underline{SU}(3)$
which can exponentiate to $S\left(U(2)\times U(3)\right)$ .

The coupling constants of a gauge theory parametrize solutions for
$g_{ab}$; e.g., for the standard model there is a three parameter
family of invariant inner products , parametrized by $\alpha_{QCD},\alpha_{QED},\theta_{W}$.

Even though the gauge group has to be compact, the fermionic matter
might have an approximate symmetry under a non-compact group $G$
that contains $K$as a subgroup. A unitary representation of $G$
will decompose into an infinite direct sum of unirreps of $K\subset G$, as we discussed earlier. 

\section{Acknowledgement}
We thank  A. P. Balachandran, G. Ferretti, D-K. Hong and K. Gupta for discussions.

\appendix
\section{The Hurwitz Zeta function and Bernoulli Polynomials}

We need to make sense of sums of the type $\sum_{\nu=0}^{\infty}e_{\nu}^{r}$
where $\nu=k+\nu$:

\[
z(r,k)=\sum_{\nu=0}^{\infty}\left(k+\nu\right)^{r}
\]

where $r$ is an even positive number. This sum is of course, divergent.
But it can be given a meaning by analytic continuation of the series
\[
\zeta_{H}(s,a)=\sum_{\nu=0}^{\infty}(a+\nu)^{-s}
\]

which converges when $\mathrm{Re\ }s>1.$ This is the well-known Hurwitz
zeta function\cite{Apostol}. It can be extended by analytic continuation
to the whole complex plane, the only singularity being a simple pole
at $s=1$. So we can define

\[
z(r,k)=\zeta(-r,k)
\]

From the standard theory\cite{Apostol} we can determine that 

\[
z(r,k)=\zeta(-r,k)=-\frac{B_{r+1}(k)}{r+1}
\]

where $B_{r}(a)$ is the Bernoulli polynomial of order $r$. There
is a useful generating function for these polynomials:

\[
B(t,a)\equiv\sum_{r=0}^{\infty}B_{r}(a)\frac{t^{r}}{r!}=\frac{te^{at}}{e^{t}-1}
\]

as well as a finite sum

\[
B_{r}(a)=\sum_{l=0}^{r}\left[\frac{1}{l+1}\sum_{m=0}^{l}(-1)^{m}\left(\begin{array}{c}
l\\
m
\end{array}\right)(a+m)^{k}\right]
\]

It is amusing to note that 

\[
z(0,k)=\frac{1}{2}-k
\]

This is the ``virtual dimension'' of the discrete series representation
of $SU(1,1)$.

The values of $z(r,k)$ for even $r$ (``Casimirs'' of the discrete
series representation of $SU(1,1)$) can now be obtained explicitly:

\[
\quad z(2,k)=\frac{1}{6}(k-1)k(2k-1)
\]

\[
\quad z(4,k)=-\frac{1}{30}(k-1)k(2k-1)(3k^{2}-3k-1)
\]

\[
z(6,k)=-\frac{1}{42}(k-1)k(2k-1)(1+3k-6k^{3}+3k^{4}),\cdots
\]

\subsection{$z\left(r,k\right)=0$ for $r=2,4,6\cdots$ and $k=\frac{1}{2},1$ }

Notice that these vanish for $k=\frac{1}{2}$ and $k=1$. Indeed this
is true for all the even values of $r$ due to a symmetry of the Bernoulli
generating function 
\[
B(t,a)=e^{-t}B(-t,1-a).
\]

This implies that 

\[
z(r,k)=(-1)^{r+1}z(r,1-k),\quad r=0,1,2,\cdots
\]

In particular, $z(r,k)$ change sign under $k\mapsto1-k$ for all
even $r$.So they must vanish for $k=\frac{1}{2}$ which is a fixed
point of the transformation $k\mapsto1-k$ . Also, $z(r,1)=0$ for
even $r$ as it is mapped to $z(r,0)$; and $z(r,0)=0$ for all $r$.

We can also verify the vanishing of these $z(r,k)$ at $k=\frac{1}{2},1$
more directly. 

If $a=\frac{1}{2}$ we can check that 
\[
\frac{te^{\frac{1}{2}t}}{e^{t}-1}=\frac{t}{e^{\frac{1}{2}t}-e^{-\frac{1}{2}t}}
\]

is an even function. So all the $B_{r}\left(\frac{1}{2}\right)$ vanish
for odd $r$. In other words 

\[
z\left(r,\frac{1}{2}\right)=0,\quad r\ \mathrm{even}
\]

Similarly when $a=1$
\[
\frac{te^{t}}{e^{t}-1}=\frac{t}{2}+\frac{1}{2}\frac{t}{e^{\frac{t}{2}}-e^{\frac{t}{2}}}\left(e^{\frac{t}{2}}+e^{\frac{t}{2}}\right)
\]

Except for the first term this is an even function. Thus $B_{r}(1)$
vanishes for all odd $r>1$. In other words,

\[
z\left(r,1\right)=0,\quad r=2,4,6,8\cdots
\]

As noted in the text, these have interesting consequences for the beta function of abelian
gauge theory.

\end{document}